\documentclass[a4paper, conference]{IEEEtran}
\IEEEoverridecommandlockouts
\usepackage{cite}
\usepackage{amsmath,amssymb,amsfonts}
\usepackage{algorithmic}
\usepackage{graphicx}
\usepackage{textcomp}
\usepackage{xcolor}
\usepackage{amsmath}
\usepackage{units}
\usepackage{alltt}
\usepackage{hyperref}

\usepackage{lipsum}
\usepackage[pscoord]{eso-pic}
\newcommand{\placetextbox}[3]{
  \setbox0=\hbox{#3}
  \AddToShipoutPictureFG*{
    \put(\LenToUnit{#1\paperwidth},\LenToUnit{#2\paperheight}){\vtop{{\null}\makebox[0pt][c]{#3}}}%
  }%
}%

\newcommand{\QRc}{QRscript}

\begin{document}
\placetextbox{0.5}{1}{This is the author's version of an article that has been published in this journal.}
\placetextbox{0.5}{0.985}{Changes were made to this version by the publisher prior to publication.}
\placetextbox{0.5}{0.97}{The final version of record is available at \href{https://doi.org/10.1109/ETFA52439.2022.9921530}{https://doi.org/10.1109/ETFA52439.2022.9921530}}%
\placetextbox{0.5}{0.05}{Copyright (c) 2024 IEEE. Personal use is permitted.}
\placetextbox{0.5}{0.035}{For any other purposes, permission must be obtained from the IEEE by emailing pubs-permissions@ieee.org.}%

\title{QRscript: Embedding a Programming Language in QR codes to support Decision and Management
}

\author{Stefano Scanzio, Gianluca Cena, and Adriano Valenzano\\
National Research Council of Italy (CNR--IEIIT), Corso Duca degli Abruzzi 24, I-10129 Torino, Italy\\
Email: \{stefano.scanzio, gianluca.cena, adriano.valenzano\}@ieiit.cnr.it\\
}

\maketitle

\begin{abstract}
Embedding a programming language in a QR code is a new and extremely promising opportunity, as it makes devices and objects smarter without necessarily requiring an Internet connection. 
In this paper, all the steps needed to translate a program written in a high-level programming language to its binary representation encoded in a QR code, 
and the opposite process that, starting from the QR code, executes it by means of a virtual machine, have been carefully detailed. 
The proposed programming language was named \QRc{}, and can be easily extended so as to integrate new features.

One of the main design goals was to produce a very compact target binary code. In particular, in this work we propose a specific sub-language (a dialect) that is aimed at encoding decision trees. Besides industrial scenarios, this is useful in many other application fields. The reported example, related to the configuration of an industrial networked device, highlights the potential of the proposed technology, 
and permits to better understand all the translation steps.
\end{abstract}


\section{Introduction}
World is changing, digital networks and connectivity are more and more pervasive, 
and communication between intelligent objects is nowadays extremely common,
to the point that this trend was explicitly given a specific name, that is,
the Internet of Things (IoT) \cite{ATZORI20102787}. 
Industry is changing too, and paradigms like Industry 4.0 \cite{CANAS2021107379} and the Industrial IoT (IIoT) \cite{8401919}, 
as well as the coexistence of heterogeneous networks \cite{SCANZIO2021103388}, 
including both wired and wireless ones, in the same production line, 
are a direct proof of this evolution.

In this scenario, configuration, operation, management, and maintenance of both new (greenfield) and old (brownfield) machinery \cite{9114523} are becoming increasingly complex tasks.
On the other hand, these essential activities must be kept as simple as possible,
is such a way to be easily accomplished by workers in the industry.
Concerning equipment involved in networked systems, 
both wired (e.g., TSN \cite{9613442}) and wireless communication technologies are often characterized by tight constraints,
including real-time \cite{8573159}, 
reliability \cite{9187609, 8060997}, 
safety \cite{9617589}, 
security \cite{CHEMINOD2019186, 8247728}, 
and power consumption \cite{10.1007/978-3-030-61746-2_11,electronics11030304}. 
Therefore, proper configuration is demanded to maximize performance.
One of the most complex parts related to the management of these kinds of apparatus is related to their diagnosis and maintenance 
in the case of malfunction \cite{MULLER20081165}. 

Many recent proposals available in the scientific literature rely on the use of portable smart devices \cite{Permin2019}, and in some case augmented reality \cite{10.1007/978-3-540-74819-9_37}, to guide the worker throughout the steps involved in maintenance operations, which must be correctly executed also by non-perfectly trained personnel.
Setting up an interactive process between the worker and an application executed on a portable device aimed, e.g., at identifying and solving a problem that has arisen in some part of the system, is an effective way to ease the implementation of the above actions, which in the most general case include configuration, operation, management, and maintenance.

Such activities typically require that the portable device is connected to either a server deployed on a local network or directly to the Internet, 
to retrieve all the required information or whatever is needed.
Sometimes, information can even be stored in the device itself, 
but doing so limits flexibility tangibly,
because one has to know in advance the kind of equipment involved and the related problems.

Unfortunately, in many cases the portable devices exploited to perform these operations do not have the possibility to use a communication network. 
This may be due to specific security policies within the company,
which prevent access to the local network,
or more commonly to the fact that the machinery is located in some place where there is no Internet access. 
Examples are installations in desert areas (high mountains, deserts, forests, etc.), 
petrochemical factories that, due to their extension, are sometimes not completely covered by a wireless network, 
railway exchanges, electrical substations, and so on.

The solution we are proposing in this work is to embed a programming language, 
we named \QRc{}, directly into a QR code,
which is attached to the related part of the equipment. 
The worker can then scan the QR code with its portable device and execute the embedded program. 
In this way, the worker is enabled to interact with the program encoded in the QR code, 
which guides her/him toward the solution of a specific problem. 
To the best of our knowledge, this solution has not been proposed before, 
and this is the first time a QR code is used to store executable code.

It is worth pointing out that the application contexts where this technology can be profitably exploited are quite wide, and
fall well outside the boundaries of factory automation. 
For example, it can be adopted in mountain trails to suggest possible routes with their respective characteristics, 
or to guide a user to correctly managing an emergency medical device, 
or finally to facilitate the use of devices of any kind in developing countries\footnote{For example, in the Suzana village in Guinea-Bissau, many residents have mobile phones, but the Internet connection is available only in some areas and for short periods of time near humanitarian associations or missions.},
by providing a certain degree of interactivity without the need for a network connection. 

In this work, in addition to presenting the main features of the \QRc{} programming language and its inherent ability to be extended, 
we also propose a specific sub-language (that is, a dialect) that permits to encode a decision tree within a QR code.
All the steps involved in translating this dialect from a high-level programming language to a binary representation 
and generating the related QR code were analyzed in depth. 
Moreover, the opposite direction of the translation scheme, 
which goes from the QR code to the execution of the encoded program, 
has been also described and commented.
Descriptions have been accompanied by a simple yet concrete example,
to make it easier to understand all the involved steps.

This paper has the following structure: 
in Section~\ref{sec:QR_techonology} QR codes are briefly described, while
the \QRc{} programming language and the main steps of the translation process 
(i.e., generation of the QR code and its execution) are defined in Section~\ref{sec:QRcoding_language}. 
The specific dialect for encoding decision trees and the related example are thoroughly detailed in Section~\ref{sec:QRcoding_DTD}. 
Finally, some conclusive remarks are reported in Section~\ref{sec:conclusions}.

\section{QR code technology}
\label{sec:QR_techonology}
The quick response code technology \cite{7966807}, commonly known as QR code, 
is a two-dimensional barcode that was invented in $1994$ with the aim of tracking vehicles during the manufacturing process. 
Besides the speed of recognition, its main advantage is the larger storage capacity when compared to pre-existing one-dimensional barcodes. 
The most recent standard specification related to QR code can be found in \cite{ISO18004}.

Currently, QR codes are used to encode different kinds of information like text, 
URLs for automatically connecting to a web page, and information to join a Wi-Fi network.
In addition, they enable different types of applications, 
for instance to manage security \cite{7569235} or mobility \cite{8825047},
for authentication, for payments, 
for augmented reality \cite{10.1145/1670252.1670305}, 
for marketing purposes \cite{marketing}, etc. 

QR codes can store four types of data, namely,
\textit{numeric}, \textit{alphanumeric}, \textit{binary}, or \textit{kanij}. 
The latter is specifically intended for encoding Japanese symbols, 
and it was included because this technology was firstly proposed by Denso Wave, 
a Japanese automotive company. 
Different data types correspond to different information that can be encoded (i.e., numeric, alphanumeric, etc.), 
and consequently their correct use may lead to a reduction (or, vice-versa, to an increase) 
of the amount of information that can be embedded in the QR code.

For the purposes of this work, we rely on the binary mode, 
which is typically used to encode 8-bit ASCII characters with the ISO 8859-1 format. 
In our case, the code included in the QR code is the result of a compilation process, 
and hence it cannot be described efficiently using the ASCII coding scheme.
Conversely, it is just a sequence of bits that represents 
the list of instructions of the program to be executed, 
which is suitably encoded in a specific binary format with the rules explained 
below.

\begin{figure}[t]
	\begin{center}
	\includegraphics[width=0.9\columnwidth]{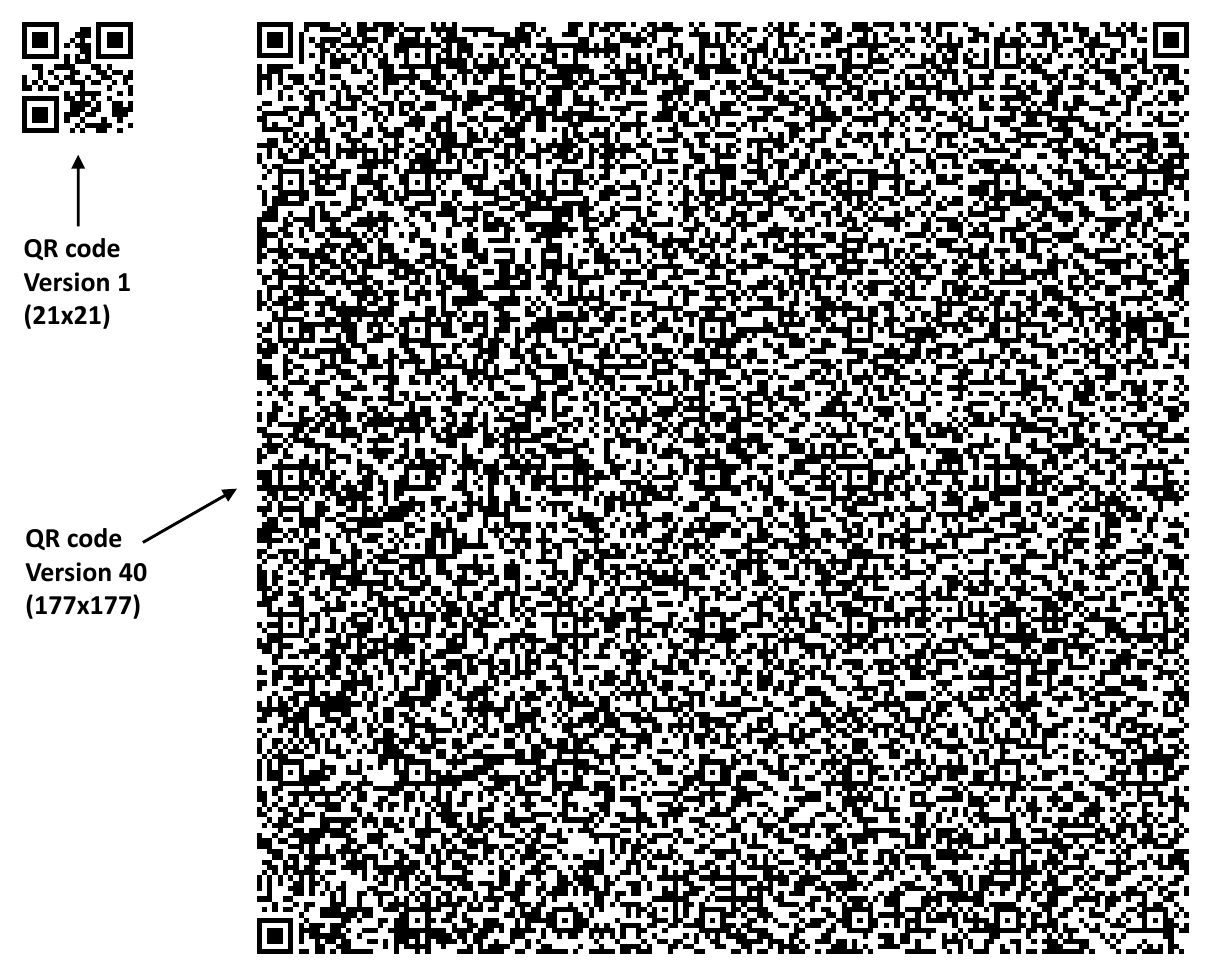}
	\end{center}
	\caption{Examples of QR code version 1 and version 40.}
	\label{fig:QRcodes}
	\vspace{0.5cm}
\end{figure}

Different \textit{versions} of QR codes are defined: 
the smallest one, in terms of the area and storing capacity is version 1, 
which is coded in a $21 \times 21$ matrix, 
while the largest is version 40, which is coded in a $177 \times 177$ matrix. 
Figure~\ref{fig:QRcodes} shows two examples of QR codes, belonging to version 1 and version 40. 
In addition, there is the possibility to define different error correction levels, i.e., L (low), M (medium), Q (quartile), and H (high), which can be employed to counteract reading issues related, e.g., to damaged or dirty QR labels, hence increasing overall robustness.
In the case of version 40 with a low level of error correction, 
the storage capacity is $2953$ bytes.

The generation of compact QR codes
and techniques to compress the information to be included 
were the subject of several scientific works. 
To improve storage capacity, colored QR codes \cite{EEI2481, 8710429} were defined, 
as well as other techniques like multiplexing \cite{6553938}.
Due to the limited storage capacity of this technology, 
the ability to pack as much information as possible in a QR code is a primary requirement in most practical applications,
and our proposal for embedding an executable program makes no exception.
For this reason great attention was spent in the translation process for generating an extremely compact binary code.

\section{\QRc{} programming language}
\label{sec:QRcoding_language}
\begin{figure*}[t]
	\begin{center}
	\includegraphics[width=2\columnwidth]{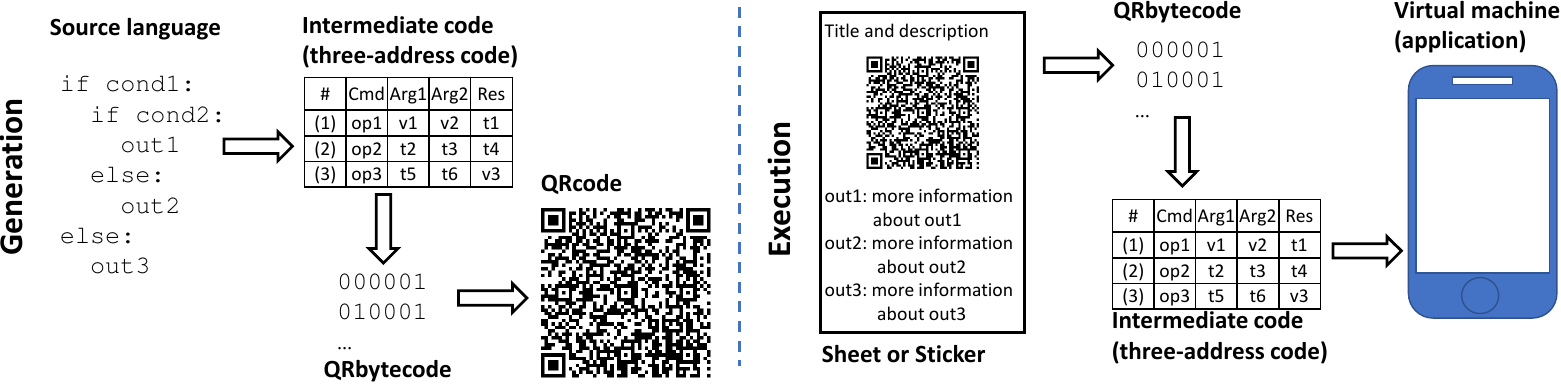}
	\end{center}
	\caption{Example of all the steps involved in the generation of the QR code (left side) and in the execution of the program from the QR code (right side).}
	\label{fig:steps}
\end{figure*}

The \QRc{} programming language is an \textit{interpreted} language that is defined inside and represented by a QR code. 
Figure~\ref{fig:steps} highlights all the steps involved in the \textit{generation} of the QR code and in the \textit{execution} of the embedded program. 
The latter step starts with the QR code being read (scanned) by the user
and ends with its execution in a specific virtual machine. 
Here, the virtual machine has to be intended as a software environment that executes the binary code retrieved from the QR code.
From a conceptual point of view, its purpose and behavior are similar to 
a Java Virtual Machine (JVM) for \texttt{java} 
or a \texttt{python} interpreter (which is actually a virtual machine).

\subsection{QRbytecode generation}
The program to be encoded in the QR code can be described by means of a high-level programming language. 
The target of this first step is the generation of a sort of \textit{bytecode}, we named \textit{QRbytecode}, 
which represents the sequence of bits that must be encoded in the QR code in a binary form. 
The reason why we defined it as a bytecode is because the program is not intended to be executed by a dedicated hardware platform, 
but rather it is interpreted by a specific virtual machine. 
The main target of the \QRc{} programming language is the definition of the QRbytecode, 
which has stringent requirements about the dimension of the generated code, 
because it has to be embedded in a QR code whose maximum size is currently $2953$ bytes 
(when version 40 with low error correction is employed).

The high-level source programming language is not the focus of this paper, 
and can be any language translatable into QRbytecode. 
Due to the aforementioned constraints related to code dimension 
and to satisfy different programming needs,
a number of \textit{dialects} (e.g., sub-languages) can be defined,
each one with different features and expressive power. 
By limiting the kinds of operations that can be performed, 
some dialects may allow the generation of a more compact QRbytecode.
At the same time, doing so constrains the characteristics and constructs 
of the high-level programming language.

The simplest dialect, which is the one we describe in this work, only supports decision trees. 
In this case, the source language could be limited to input operations, output operations, and ``\texttt{if}'' statements. 
Consequently, only the sequence and selection (choice) constructs of structured programming languages can be implemented, 
but not the repetition (loop). 
In addition, the ability to define and use variables could be omitted in some specific dialects. 
This implies that for these dialects the source programming language is necessarily streamlined, 
and only a small subset of algorithms can be actually implemented
(which, however, includes many of those we are actually interested in). 

As said before, the exact definition of the high-level programming language is outside the scope of \QRc{}, 
therefore all the pieces of code written in this language that appear in this work must be considered 
only as examples of a possible source language.
When performing the translation from the source programming language to the QRbytecode, 
an intermediate representation can be possibly employed, like the \textit{three-address code}. 
Typically, doing so eases both the design and the implementation of the translator. 
For this reason, in this work it was decided to follow this direction.

Once the QRbytecode has been generated and possibly optimized, 
it can be encoded in a QR code and, eventually, printed on a sheet or sticker that can be attached to the related industrial machinery, in a convenient place to be seen and used by workers. 
In addition to the QR code itself, the sticker can also contain some printed information
written in natural language,
which explains how to use the QR code and how to interpret the output of the program.

\subsection{QRbytecode execution}
The execution process starts when the QR code is read by a client device 
(a smartphone, a tablet, or more in general an application) that,
as the first step, performs the conversion from the QR code back to the QRbytecode. 
Then, the interpreter transforms this language into a suitable internal representation 
(which is typically the tree-address code) and executes it by means of a dedicated virtual machine.

From a practical point of view,
all the applications installed in the client need 
the ability to interpret and execute the code stored in the QRbytecode. 
Typically, it is an Android or Apple iOS app, 
which can be possibly customized for specific application contexts 
and modified by the system manager (or users, in general) 
so as to meet some requirements. 
In particular, the way input and output operations are performed is highly dependent on the characteristics of the industrial process and environment. 
For instance, in dusty/dirty industrial environments, 
input can be made easier if the application uses large fonts for the condition 
(i.e., when asking questions) 
and large on-screen buttons for every possible response. 
Likewise, in contexts where hands-free operation is demanded,
the use of text to speech (TTS) and automatic speech recognition (ASR) modules 
could facilitate the interaction with the worker.

It is worth stressing that QR codes embedding QRscript programs are mostly valuable when the device used to carry out configuration and guided diagnostics is not provided with a reliable/stable Internet connection, otherwise a simple web browser 
coupled with a QR code containing the URL of a suitable web resource is perhaps the best choice.
This may also happen in those installations where wireless access is not authorized 
for devices of the maintenance staff.

\subsection{\QRc{} dialects}

\begin{figure}[t]
	\begin{center}
	\includegraphics[width=0.9\columnwidth]{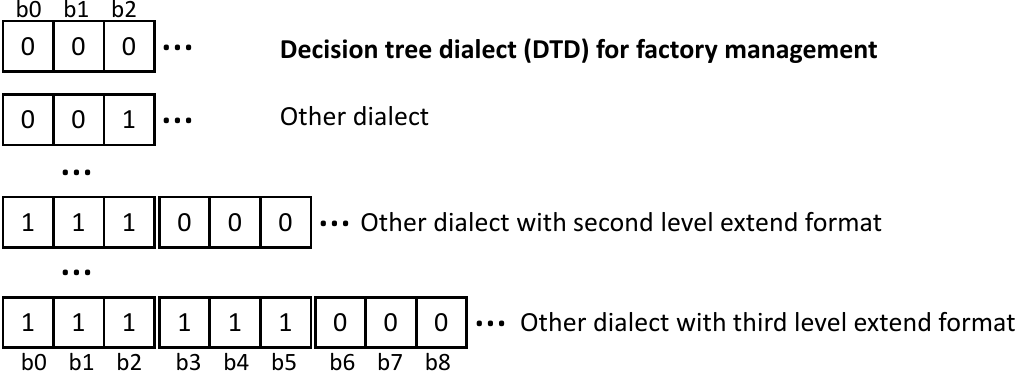}
	\end{center}
	\caption{Definition of the initial part of the QRbytecode specifying the dialect.}
	\label{fig:dialects}
	\vspace{0.5cm}
\end{figure}

Different dialects can be defined for the \QRc{} programming language, 
in order to reach a compromise between the dimension of the generated QRbytecode 
and the expressiveness of the programming language. In the current proposal,
the first $3$ bits of the QRbytecode are used to identify the dialect that is being used in the following part of the code. 
As shown in the schema of Figure~\ref{fig:dialects}, 
the dialect characterized by code \texttt{000} 
corresponds to the decision tree dialect (DTD), 
which is the one analyzed and described in this work.

\begin{figure}[t]
	\begin{center}
	\includegraphics[width=0.8\columnwidth]{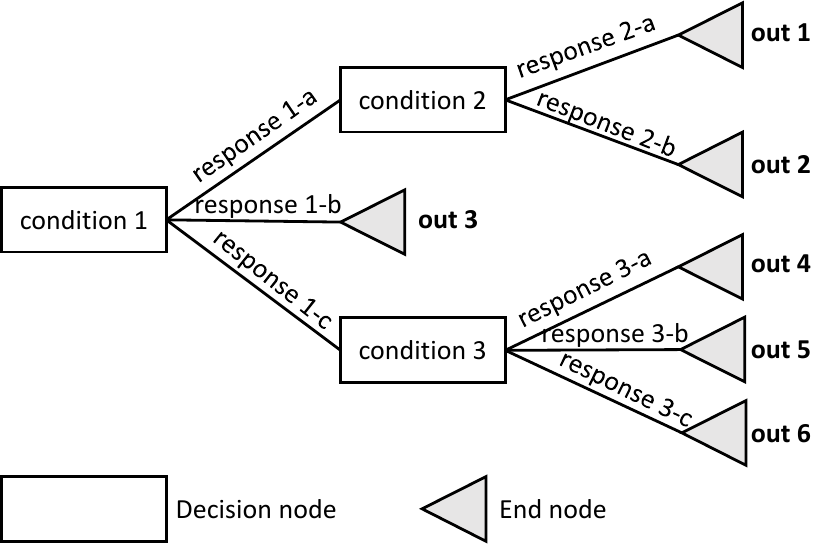}
	\end{center}
	\caption{Example of a decision tree without chance nodes.}
	\label{fig:decision_tree}
	\vspace{0.5cm}
\end{figure}

As a matter of fact, 
the operations that are requested for configuring and maintaining industrial equipment can be often described using a simplified version of a \textit{decision tree} model, with only \textit{decision} and \textit{end} nodes, but without any \textit{chance} nodes (chance nodes are a kind of nodes that represent a probabilistic decision). 
In the context of this work, we want that all the decisions are taken by the user 
who is interacting with the algorithm.
An example of this kind of decision tree is reported in Figure~\ref{fig:decision_tree}, 
in which every decision node can have more than two responses 
(we decided not to limit it to binary decisions). 

For every outcome, which is represented by an end node, 
a specific \textit{decision rule} is associated. 
In the case of the decision tree of Figure~\ref{fig:decision_tree}, 
six decision rules can be generated.
For the first outcome the decision rule is 
\texttt{if (condition 1 == response 1-a) AND (condition 2 == response 2-a) then out 1}. 
A decision tree can be directly (and easily) transformed into a \textit{flow chart}, 
which is a more efficient representation because it does not require the use of variables to store the results of conditions, 
and is also more deterministic and faster from the point of view of the execution time.

The other combinations of the first 3 bits of the QRbytecode can be used to define other dialects. 
As previously mentioned, 
they are just languages with different capabilities in terms of the available instruction set and the dimension of the QRbytecode. 
As far as we know, this is the first time that a programming language is compiled and fit into a QR code. 
Nevertheless, in the case some pre-existing languages were defined, they could be easily included 
in this representation by assigning them a specific dialect code. 
The format for representing dialects is defined in such a way that it can be easily extended to a number of dialects that exceeds what can be encoded on three bits. 
In particular, if the first three bits are equal to \texttt{111}, the next $3$ bits are used to identify additional dialects.
Likewise, if the first six bits are all equal to \texttt{1}, the next $3$ bits can be used to identify new dialects, and so on.

\section{Decision tree dialect}
\label{sec:QRcoding_DTD}
The DTD is a dialect of the \QRc{} programming language 
that permits to encode well-known decision trees without chance nodes.
Consequently, it can be exploited to define algorithms to help workers to solve certain problems, 
which are for instance related to network (or machinery) configuration and maintenance.
In DTD, an improved version of decision tree is defined that supports the concatenation of several decision trees. 
In particular, every condition has a default response, we named \textit{other}, 
that, if selected, permits the user to move to the following decision tree.

In this subsection, DTD is described and defined starting from a compact and illustrative example. 
Some of the main parts involved in the generation of the QRbytecode,
starting from a sample high-level programming language, 
were implemented using the \texttt{jflex} scanner and the \texttt{cup} parser, 
which produce a translator implemented in the \texttt{java} language.
However, they can be implemented with any other bottom-up parser, 
like the pair \texttt{flex/bison} that produces a translator in the \texttt{C} programming language, 
or \texttt{ply} that produces a translator in the \texttt{Python} programming language. 

To make understanding easier, we explained this dialect with an example in a top-down fashion, 
i.e., starting from a high-level graphical representation to arrive to the binary QRbytecode that is encoded in the QR code.

\subsection{High-level representation}
\begin{figure}[t]
	\begin{center}
	\includegraphics[width=\columnwidth]{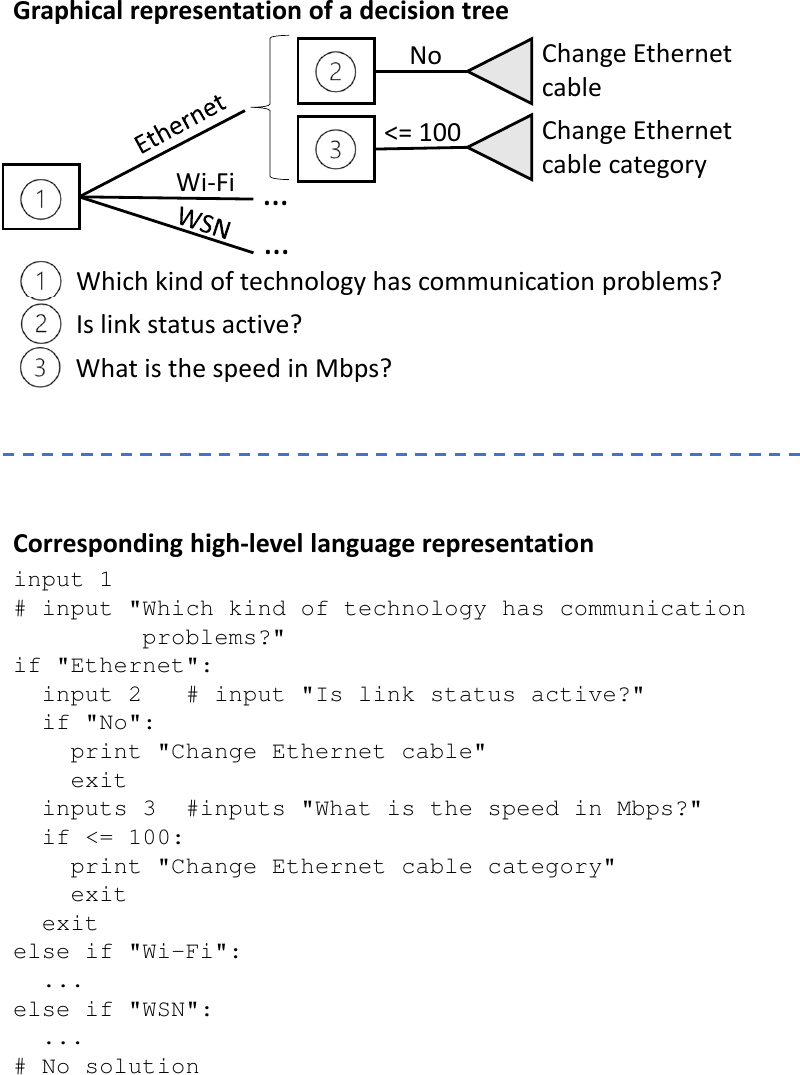}
	\end{center}
	\caption{Example of decision tree and corresponding high-level language that can be encoded in a QR code using \QRc{} (``\#'' denotes comments).}
	\label{fig:high_level}
\end{figure}

Figure~\ref{fig:high_level} shows the graphical representation of a decision tree that can be implemented using \QRc{}, and the corresponding representation through a Python-like pseudo-code. 
In this simplified example, the decision tree is used to guide the worker in solving a connectivity problem. 
After asking the user to identify the kind of technology that is having a problem (i.e., Ethernet, Wi-Fi, or WSN), 
in the case the response is ``Ethernet'' the algorithm described by the decision tree tries to identify the problem by asking if the link status is active. 
Typically, if the Ethernet cable is disconnected or broken the link status led is inactive. 
If the user selects a response other than ``No'',
the algorithm switches to the next concatenated decision tree, 
identified with the number \raisebox{.5pt}{\textcircled{\raisebox{-.9pt} {3}}} as root, which corresponds to the question ``What is the speed in Mbps?''. 
If the value entered by the user does not match the condition ``\texttt{<= 100}'' 
the algorithm ends without a solution, otherwise it recommends to ``Change Ethernet cable category''
(we assume that cabling must support Gigabit Ethernet, whose expected speed is $\unit[1000]{Mb/s}$).

The representation of the algorithm with a high-level programming language is reported in the lower part of the same figure. 
The \texttt{input/inputs} instruction prints a message on the screen 
and requests a string as input, 
which can be entered by the user either \textit{directly} through a textual interface,
when the \texttt{inputs} instruction is used, 
or \textit{indirectly}, for instance by means of a decision button,
in the case of the \texttt{input} instruction. 
In this dialect only the \texttt{string} type can be used for input values, 
and the virtual machine aimed at executing the code tracks only the last entered value (i.e., every time a new \texttt{input} instruction is executed
the previously entered string is definitely lost).

When the argument of the instruction is an unsigned integer,
as for \texttt{input 1} and \texttt{inputs 1}, 
a reference is printed instead of a string.
See, for instance, the symbol \raisebox{.5pt}{\textcircled{\raisebox{-.9pt} {1}}}. %
The idea behind this option is that the string associated to reference \raisebox{.5pt}{\textcircled{\raisebox{-.9pt} {1}}} 
can be physically printed on a sheet or a sticker
(likely the same on which the QR code is found),
instead of being encoded in QRbytecode. 
In this way, a fair amount of space is saved in the QR code, 
since the characters that make up the string are left out of it.

Comparison with integer (e.g., \texttt{<= 100}) and floating point (e.g., \texttt{> 3.5}) values is also possible. 
In these cases, the virtual machine performs an automatic conversion of the  string entered by the user into an internal integer or floating point value, respectively.

\begin{figure}[t]
	\begin{center}
	\includegraphics[width=0.65\columnwidth]{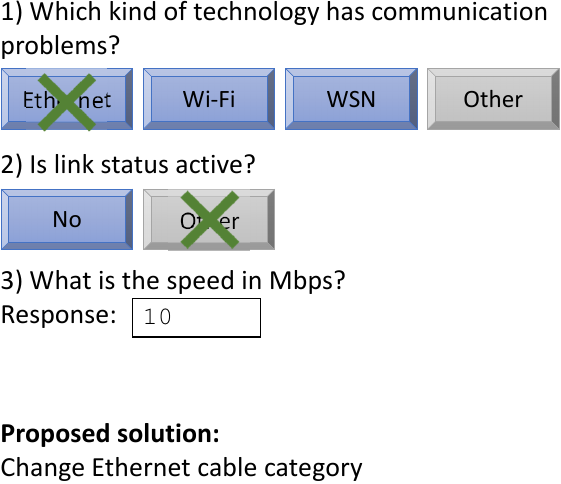}
	\end{center}
	\caption{Example of interaction between the user and the QR code mediated by a suitable application provided with a graphical interface
	(the interface can be automatically built when executing the QRbytecode of the DTD).}
	\label{fig:application}
\end{figure}

A possible interaction between a user and the application 
(i.e., the virtual machine that executes the QR code), 
based on the above discussed program, is depicted in Figure~\ref{fig:application}. 
As shown on the right side of Figure~\ref{fig:steps}, 
the application and its virtual machine 
are the final step in the chain related to code execution, 
which is directly based on the QRbytecode automatically generated by the translator and read from the QR code. 
We decided to start our discussion about the whole chain from the end, 
and to begin with the interaction with the user, 
to permit a complete and thorough understanding of the example. 
As can be seen, the questions asked by the algorithm depend on the previous responses provided by the user. 
To each question, a default response named ``Other'' (and reported in grey in the figure) is added. 
If selected, this response causes the program flow to jump to the next concatenated question (or decision tree).
In the case of input sets made up of enumerated values (e.g., related to an \texttt{input} instruction), they could be represented by a number of buttons. 
Instead, a typical rendering for an \texttt{inputs} instruction is through a textual box.

\subsection{Translation of DTD to three-address code}
\begin{figure}[]

\begin{center}
\footnotesize
\begin{alltt}
(1)  input "Which kind of technology has
                communication problems?"
(2)  if "Ethernet" (6)
(3)  if "Wi-Fi" (15)
(4)  if "WSN" (20)
(5)  goto (25)
(6)  input "Is link status active?"
(7)  if "No" (9)
(8)  goto (10)
(9)  printex "Change Ethernet cable"
(10) inputs "What is the speed in Mbps?"
(11) ifc <= 100 (13)
(12) goto (14)
(13) printex "Change Ethernet cable category"
(14) printex ""
(15) # Code related to Wi-Fi
     ...
(20) # Code related to WSN
     ...
(25) printex ""
\end{alltt}
\end{center}
  \caption{Example of three-address code derived from the high-level language representation reported in Figure~\ref{fig:high_level}.}
  \label{fig:three_address_code}
\end{figure}

Starting from a high-level description of the decision tree (graphical or textual),
an intermediate representation based on the three-address code can be automatically obtained by means of a translator.
It is worth remarking again that the high-level language proposed in this work is just an example. 
The specifications related to \QRc{} in general, and DTD in particular, 
are related to QRbytecode generation, 
and from a certain perspective to three-address code generation as well, 
because there is a direct mapping between these two formalisms.

Three-address code typically consists of a numbered/ordered list of quadruples, 
where the fields of each quadruple represent the instruction (or operation), 
its first and second arguments, and the result, respectively. 
For some instructions the second argument is not used.

\begin{figure*}[t]
	\begin{center}
	\includegraphics[width=1.9\columnwidth]{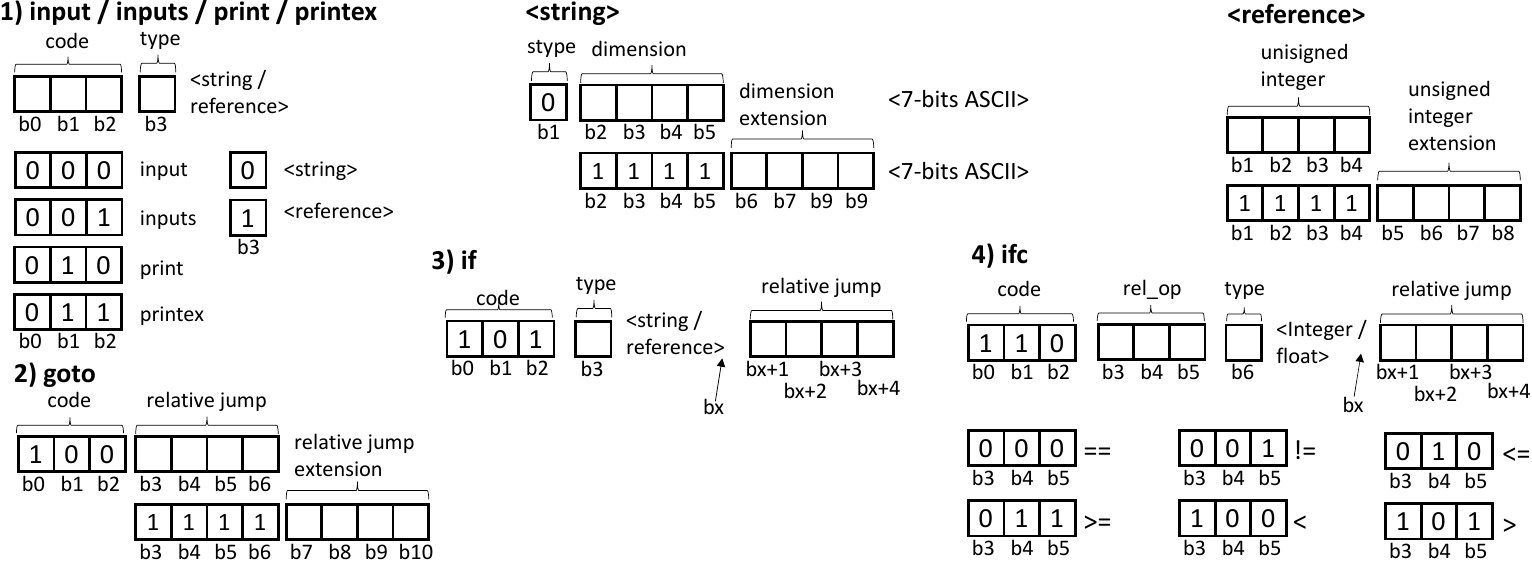}
	\end{center}
	\caption{Conversion rules from three-address code to QRbytecode.}
	\label{fig:QRbytecode}
	\vspace{-0.2cm}
\end{figure*}

In DTD three-address code, seven instructions are defined:
\begin{itemize}
    \item \texttt{input <constant>} (or \texttt{inputs <constant>}): requires an indirect (or direct) input of a string, respectively. The term \texttt{<constant>} can be either a \texttt{<string>} or a \texttt{<reference>} (i.e., an unsigned integer value).
    \item \texttt{print <constant>}: prints the string or the reference identified by \texttt{<constant>}.
    \item \texttt{printex <constant>}: same as \texttt{print}, but after the instruction the execution is terminated.
    \item \texttt{goto (x)}: jumps to instruction number \texttt{x}, which is identified by the label \texttt{(x)}.
    \item \texttt{if <string> (x)}: if the last entered string value is equal to \texttt{string}, it jumps to instruction number \texttt{x}.
    \item \texttt{ifc <rel\_op> <operand> (x)}: if the comparison between the last entered string value and the \texttt{<operand>} using the \texttt{rel\_op} relational operator is true, it jumps to instruction number \texttt{x}. If \texttt{<operand>} is an integer or a floating point value, the last entered string is automatically converted to an integer or a float number, respectively. An error is returned if the conversion is not possible. 
    Possible \texttt{rel\_op} are \texttt{==}, \texttt{!=}, \texttt{<=}, \texttt{>=}, \texttt{<}, and \texttt{>}. 
    This is the only instruction that makes use of both the arguments of the three-address code representation based on quadruples.
\end{itemize}

The most important and complex part to pay attention in three-address code generation is related to semantic actions aimed to the generation of unconditional (\texttt{goto} instruction) and conditional (\texttt{if} and \texttt{ifc} instructions) jumps. 
In any case, this aspect is completely addressed by compilers' theory, 
and for bottom-up parsers it can be easily solved through the use of specific inherited attributes.

Figure~\ref{fig:three_address_code} reports the three-address code obtained from the decision tree described by the high-level language in Figure~\ref{fig:high_level}. 
The fields of each quadruple are separated by means of white spaces.
Analyzing one of the possible execution flows, 
if the answer to the first choice about the network technology is ``Ethernet''
a jump is performed to instruction \texttt{(6)}, 
in which the second input is requested. 
If the user, at this point, replies ``No'', 
there is a jump to instruction \texttt{(9)}.
Then, the proposed solution is consequently displayed (``Change Ethernet cable''), after which the program terminates. 
In the case the reply to the second question in line \texttt{(6)} is ``Other''
(whose meaning is that link status is active), 
a jump is made to instruction \texttt{(10)}, 
and another input is requested, which in this case is direct. 
If the user enters a value greater than $100$, 
the \texttt{goto (14)} instruction is executed.
In this case the program jumps to line \texttt{(14)} and, with the instruction \texttt{printex ""}, it terminates its execution without printing any message. 

\subsection{QRbytecode generation for DTD}
Starting from the three-address code representation, the QRbytecode of the DTD can be directly generated. 
In this step, the main goal we pursued was to reduce the dimension of the generated code as much as possible.
Consequently, when defining the translation rules, we paid great attention to generating a compact code. 
This section is aimed at the definition of the syntax (and semantics) of the QRbytecode, 
and it shows how the DTD dialect can be transformed bitwise
into bytecode and how, starting from bytecode, 
the relevant three-address code can be retrieved. 
The description on how to implement the virtual machine that executes the QRbytecode has not been reported because it does not add any relevant details. 

The first three bits of the QRbytecode are \texttt{000}, 
to identify the DTD dialect and to instruct the virtual machine to use DTD rules. 
Each instruction in three-address code has a direct mapping in QRbytecode.
In particular, the first three bits (i.e., the instruction \textit{code}) 
identify the instruction: 
\texttt{000} for \texttt{input}, 
\texttt{001} for \texttt{inputs}, 
\texttt{010} for \texttt{print}, 
\texttt{011} for \texttt{printex}, 
\texttt{100} for \texttt{goto}, 
\texttt{101} for \texttt{if}, and 
\texttt{110} for \texttt{ifc}. 
The code \texttt{111} was left unused to permit possible extension of the instruction set. 
A schematic view of how the instructions have been encoded is reported in Figure~\ref{fig:QRbytecode}.

Starting from the input/output instructions 
(i.e., \texttt{input}, \texttt{inputs}, \texttt{print}, and \texttt{printex}), 
the \textit{type} bit after the code identifies whether the provided output 
is either a string (\texttt{<string>}) or a reference (\texttt{<reference>}).

Since storing strings requires a fair amount of space, we paid attention to their definition. 
Each string in its binary representation starts with a \textit{stype} bit which, if set to \texttt{0}, defines a compact string, 
and it is followed by a \textit{dimension}, encoded as an unsigned integer on $4$ bits,
which represents the number of characters that are subsequently encoded. 
Similarly to the encoding of dialects, 
when the previous $4$ bits are equal to \texttt{1111}, the dimension is extended recursively by another $4$ bits. 
Characters are encoded using a 7-bit character set, typically ASCII, 
but if both the generation and the execution steps agree on the standard to be used, 
other solutions like ISO/IEC 646 and 8-bit character sets are other viable options. 
Clearly, the latter increases the dimension of the generate binary code.
The value \textit{stype=1} was currently left unused (reserved), 
and can be exploited to extend string coding to other character sets.
References are coded as unsigned integers on $4$ bits, 
and can be recursively extended to represent numbers of any size.

The \texttt{goto} instruction executes unconditional jumps, 
and is identified by the starting sequence of bits \texttt{100}. 
The \textit{code} is followed by a \textit{relative jump}, 
which specifies a recursively extensible jump width,
encoded as an unsigned integer on $4$ bits, 
that represents the number of instructions to be skipped starting from the next instruction 
(i.e., \texttt{0000} means jump to the next instruction). 
It is worth pointing out that only forward jumps are possible, because DTD does not foresee loops. 
Conversely, in the case of a dialect with backward jumps, 
the relative jump has to be encoded as a signed integer.

Two conditional jump instructions are defined, 
namely \texttt{if} and \texttt{ifc}. 
The first is identified by the code \texttt{101}, 
followed by a type that specifies whether the value inserted by the user has to be compared with either a string or a reference, 
and a relative jump is executed when the comparison matches.
The \texttt{ifc} instruction differs from \texttt{if} because, 
after the code \texttt{110}, 
it defines a relation operator (\textit{rel\_op)} using a $3$ bits representation. %
For this instruction, \textit{type=0} identifies a signed integer number, 
while \textit{type=1} identifies real number encoded as a half-precision floating point on $16$ bits. 
Integer numbers are stored using the \textit{two's complement} representation with $16$ and $32$ bits. The first bit is used to select between $16$ and $32$ bits representations.

At the end of the translation process, 
the QRbytecode is padded with up to $7$ bits taken from the sequence \texttt{1000000}, in such a way that the final dimension of the generated QRbytecode is a multiple of $8$ bits, 
as possibly requested by the \textit{binary} coding mode. 
If the number of bits added by padding is not enough to encode a complete instruction,
the virtual machine will simply discard it at runtime.
Otherwise, the instruction the virtual machine will interpret 
in the padding corresponds to a \texttt{goto} conditional jump to the next instruction (that does not exists).
In both cases no problems are caused to the virtual machine. 

Using the translation rules described above, 
the three-address code presented in Figure~\ref{fig:three_address_code} 
can be directly translated to QRbytecode, and vice-versa. In terms of storage occupation, the space needed to store instructions from (1) to (14) of Figure~\ref{fig:three_address_code} is $654$ bits ($82$ bytes with padding).
Instead, the translation between QRbytecode and QR code is trivial, 
because a number of libraries exist free of charge to perform this operation\footnote{An example is the \texttt{PyQRCode} module for the \texttt{Python} programming language.}.

\section{Conclusions}
\label{sec:conclusions}
The ability to embed an executable program inside a QR code is of particular interest, because it enables a number of applications. 
Decision support and configuration, 
as well as maintenance and diagnostics of industrial equipment and networks, 
are just few examples of the possible scenarios that can take advantage by this technology. 
Systems located far away from the main industrial plant, 
where Internet access is not available (or unreliable) for whatever reason, 
including military applications, can all benefit from our technology.
Unlike techniques that make the device used for configuration (a mobile phone) interact directly
with intelligent equipment by means of, e.g., NFC or visible light communication,
our solution does not require any modification to the existing equipment and is extremely inexpensive.

The proposed \QRc{} programming language has been conceived bearing extensibility in mind, 
and permits to define a number of sub-languages termed dialects. 
In this work, all the steps needed to generate a QR code containing a program in binary form 
starting from its description given in a high-level programming language,
and the corresponding chain with which the program is read from the QR code and executed on a virtual machine, 
have been thoroughly detailed. 
The DTD dialect, which is aimed at coding a decision tree inside a QR code, was carefully defined,
along with an intermediate representation of the program based on three-address code. 
The description of the dialect and the translation process were explained by using a very simple, yet realistic example. 

We believe that the ability to embed a program in a QR code that can be read and executed on handheld/portable devices is a really appealing option.
Consequently, our future works on this subject will focus on identifying new areas where this technology can be profitably employed, and on providing concrete examples to show how this can be done.
Moreover, we plan to define additional dialects in order to enhance the features and capabilities of \QRc{}, and to add security features based on asymmetric cryptography to distinguish  among user roles (those enabled to forge QR codes, to just execute them, or to do nothing).

\bibliographystyle{IEEEtran}
\bibliography{bibliography}

\end{document}